\documentclass[conference]{IEEEtran}
\IEEEoverridecommandlockouts
\usepackage{cite}
\usepackage{amsmath,amssymb,amsfonts}
\usepackage{algorithmic}
\usepackage{graphicx}
\usepackage{textcomp}
\usepackage{xcolor}
\usepackage{multirow}
\usepackage{url}
\usepackage[T1]{fontenc}
\usepackage{booktabs}
\usepackage{comment}
\usepackage{wrapfig}

\newcommand{\joaquin}[1]{{\color{olive}{\textbf{Joaquin}: #1}}}

\def\BibTeX{{\rm B\kern-.05em{\sc i\kern-.025em b}\kern-.08em
    T\kern-.1667em\lower.7ex\hbox{E}\kern-.125emX}}
\begin{document}

\title{Adaptive, Continuous Entanglement Generation for Quantum Networks\\
\thanks{A. K. and A. Z. contributed equally to this work. This material is partly based upon work supported by the U.S. Department of Energy, Office of Science, National Quantum Information Science Research Centers.}
}

\author{\IEEEauthorblockN{Alexander Kolar, Allen Zang}
\IEEEauthorblockA{
\textit{University of Chicago}\\
atkolar@uchicago.edu, yzang@uchicago.edu}
\and
\IEEEauthorblockN{Joaquin Chung, Martin Suchara, Rajkumar Kettimuthu}
\IEEEauthorblockA{
\textit{Argonne National Laboratory}\\
chungmiranda@anl.gov, msuchara@anl.gov, kettimut@mcs.anl.gov}
}


\maketitle

\begin{abstract}
Quantum networks, which enable the transfer of quantum information across long distances, promise to provide exciting benefits and new possibilities in many areas including communication, computation, security, and metrology. These networks rely on entanglement between qubits at distant nodes to transmit information; however, creation of these \textit{quantum links} is not dependent on the information to be transmitted. Researchers have explored schemes for continuous generation of entanglement, where network nodes may generate entanglement links before receiving user requests. In this paper we present an adaptive scheme that uses information from previous requests to better guide the choice of randomly generated quantum links before future requests are received. We analyze parameter spaces where such a scheme may provide benefit and observe an increase
in performance of up to 75\% over other continuous schemes on single-bottleneck and autonomous systems networks.
We also test the scheme for other parameter choices and observe continued benefits of up to 95\%.
The power of our adaptive scheme on a randomized request queue is demonstrated on a single-bottleneck topology. 
We also explore quantum memory allocation scenarios, where a difference in latency performance implies the necessity of optimal allocation of resources for quantum networks.
\end{abstract}

\begin{IEEEkeywords}
quantum networks, entanglement generation, quantum teleportation, adaptive protocol
\end{IEEEkeywords}

\section{Introduction}
\label{sec:intro}

Quantum information science has started another revolution of quantum computation and communication.
\emph{Quantum networks}~\cite{kimble2008quantum} which connect quantum devices and interconnect themselves are expected to become a reality, but their construction is difficult due to fundamental physical differences compared to a classical network.
Direct transmission of qubits, which encode quantum information, endures inevitable channel losses and environmental noise.
Even worse, the \emph{no-cloning theorem} 
forbids creation of identical copies of an arbitrary unknown quantum state, which means that no amplification is possible for a quantum signal.
Fortunately, \emph{Quantum teleportation} 
based on maximal \emph{entanglement} between two parties can transfer an arbitrary quantum state from one party to the other.
Entanglement thus plays a role analogous to links in classical networks.
Moreover, established entanglement links can be extended using \emph{entanglement swapping}~\cite{swapping}, where multiple shorter-distance entanglement links are consumed to produce a single long-range link.
Usage of quantum teleportation for transmission thus makes entanglement generation between nodes a prerequisite for a functional quantum network.
However, differences exist from classical links.
Entanglement is consumed after each teleportation, which makes it a one-time-use link. 
Additionally, \emph{quantum memories} storing entangled qubits have finite (and generally short) coherence times which determine the timescale of entanglement degradation.
As a result, entanglement links are strongly time-sensitive.

Given the complexity of potential realistic situations and the limitations imposed on quantum networks, including arbitrary entanglement requests and the probabilistic nature of near-term entanglement generation methods,
establishing entanglement between arbitrary node pairs in a timely manner is difficult but important. This problem is two-fold: (1) local entanglement links must be generated between intermediate nodes and (2) an optimal route between the nodes in the pair must then be found using these links.
The routing problem (2), known as \emph{entanglement routing}, is an emerging and active research topic in the quantum network community with various existent research papers on entanglement routing protocols and algorithms. 
However, in this paper we focus on the entanglement generation problem (1). Specifically, we explore continuous generation, where entanglement links are generated regardless of requests for entanglement establishment and which may reduce latencies for serving such requests. We propose a heuristic adaptive scheme for continuous entanglement generation. For entanglement routing, we adapt the proposed local best effort routing algorithm from~\cite{chakraborty2019distributed}.

The structure of this paper is as follows. In Section~\ref{sec:methodology}, we introduce our methodology for quantum network simulation, including assumptions, parameters, and explanation of our simulation procedure. In Section~\ref{sec:setup}, we explain the mechanism of our adaptive entanglement generation protocol and the setup of our simulated networks. In Section~\ref{sec:results}, we present our simulation results with corresponding discussion. In Section~\ref{sec:conclusion}, we conclude our work.

\section{Methodology}
\label{sec:methodology}

\subsection{Network Assumptions}
\label{sec:assumptions}
The simulated networks operate as simplified entanglement distribution networks. To manage entanglement, a node may choose at each simulation time step to attempt generation of a new entanglement link with an adjacent node or extend entanglement by attempting entanglement swapping.

We make the following simulation assumptions:
\begin{itemize}
    \item The adaptive, continuous entanglement generation protocol only has knowledge of adjacent nodes.
    \item Each entanglement generation attempt between adjacent nodes (and swapping between three nodes) occurs within one simulation time step.
    \item Entanglement generation and swapping have a fixed (and distinct) probability of success per attempt. 
    \item Each node contains a fixed number of quantum memories.
    \item Quantum memories have a fixed storage time, after which stored entanglement is lost and memories are automatically reset.
    For real quantum memories, the initialization (preparation) time is finite and technology-dependent. For example, the preparation time of atomic frequency comb (AFC)-type absorptive memories is on the order of 100 ms~\cite{lago2021telecom}, while could be much shorter for single-qubit memories (e.g. below 100 \textmu s~\cite{9492859,bourassa2020entanglement}).
    \item Entanglement fidelity during generation and swapping is ignored. 
    In realistic scenarios, imperfect Bell states might be generated and entanglement purification is needed to increase entanglement fidelity. However, we note that for protocols like Barrett-Kok for entanglement generation between two ions the entanglement fidelity can be very high for high Purcell factor and small detuning between the two ions ~\cite{wein2020analyzing,asadi2020protocols}.
    \item Classical communication and computation are considered ``free'' operations, as the development of classical networks is much more advanced than quantum networks.
    Issues such as classical network congestion could in principle be ignored if we consider that a classical network topologically identical to the quantum network is also implemented, forming a ``dual network''~\cite{rohde2021quantum}. Moreover, since the classical information communicated is always simple, we ignore processing time.
    Therefore, the two major quantum operations we consider (entanglement generation and swapping) share the same scale
    of operation time (the sum of on-node operation time and qubit transmission time over optical fibers). We then take this characteristic time scale as our simulation time step.
    \item A centralized scheduler schedules every request and ensures that only one request is serviced in the network at any given time.
\end{itemize}

\subsection{Simulation Parameters}
\label{sec:params}
Based on our network assumptions, we identify the following network parameters (with values discussed in~\ref{sec:net_setup}): number of nodes $N$, 
number of memories per node $m$, memory lifetime $\tau_m$, entanglement generation success probability $p_e$, and entanglement swapping success probability $p_s$. We also maintain an adjacency matrix describing the connection of nodes via optical fibers for transmitting photonic qubits. From this, we derive the degree of each node $n$, denoted as $d_{n}$.

\subsubsection{Parameter Space}
\label{sec:param_space}
Here, we qualitatively analyze which parameter regimes our adaptive protocol might be helpful for. First, we consider a simple protocol -- which is also the starting phase of our adaptive protocol as described in Section~\ref{sec:adaptive}. At each time step, a node $n$ 
chooses a neighbor node $k$ 
at random with uniform probability to attempt entanglement generation. We then have the expected number of links formed between neighboring nodes $n$ and $k$ in one time step 
\begin{equation}
    P(L) = p_e\left(\frac{1}{d_n} + \frac{1}{d_k}\right) = p_e\frac{d_n + d_k}{d_nd_k}
\end{equation}
We also compute the expected number of links $E(L)$ that node $n$ creates with all nodes in its neighbor set $K$. This is equivalent to the probability of any neighbor in $K$ randomly selecting $n$ and succeeding in generating entanglement, or that node $n$ randomly selects a node in $K$ and succeeds in generating entanglement. For our simple model, the probability of selecting a node in $K$ is one; we may then write the expected number of links as
\begin{equation}
    E(L) = p_e + \sum_{k}\frac{p_e}{d_k} = p_e\left(1 + \sum_{k}\frac{1}{d_k}\right)
\end{equation}
This number induces a time scale $1/E(L)$, after which we expect one memory will be occupied. Finally, we consider the expected number of links generated with a specific neighboring node $k\in K$ after some number of time steps $\Delta T$. If we assume that the lifetime $\tau_m$ is longer than the time interval $\Delta T$, then the number of links shared with node $k$ is bounded by the number of memories $m$ under the assumption that one quantum memory can only be entangled with one other memory (according to monogamy of entanglement\cite{monogamy}). Combining all results, we have that
\begin{equation}
    \label{eq:EL_k}
    E(L_k) = p_e\frac{d_n + d_k}{d_nd_k}\,\text{min}\left(\Delta T,\frac{m}{E(L)}\right)
\end{equation}

For parameter regimes where $E(L_k)$ is small it is more likely that certain needed entanglement links may not have been generated by a protocol with uniform probabilities before a request is scheduled. Therefore we expect our adaptive protocol to be useful in scenarios with small $E(L_k)$, i.e. with
\begin{itemize}
    \item Low entanglement generation probability $p_e$,
    \item Low number of memories $m$ with respect to node degree, 
    \item High degree (connectivity) of nodes $d_n$, or
    \item Low time interval between requests $\Delta T$.
\end{itemize}
In this paper, we focus on the first two items, which correspond to realistic expectations for near-term quantum networks. 
Considering the high cost of building and maintaining quantum memories, the number of memories per node will not be high in the near future. 
We also study small-scale networks (discussed in~\ref{sec:net_setup}) with modest average node degrees.

\subsection{Simulation Procedure} \label{sec:sim-procedure}
\label{sec:procedure}
We use a traffic matrix $T$ to represent expected average network usage. From this matrix, we generate a queue of $r$ requests $R(a,b)$ for an entanglement link between two arbitrary network nodes $a$ and $b$. Each element of the queue is chosen randomly with a distribution determined by the traffic in the traffic matrix, such that heavily trafficked connections (with a larger $T_{ab}\in[0,1]$) have a higher number of corresponding requests in the queue on average.

At the beginning of each simulation trial, time is allocated for initial entanglement generation. For our simulations, this period is equal to the interval between successive requests with a uniform request interval.
Then the first request from the queue is passed to the network. 
After a request is completed (an entanglement link has been established between the two nodes), 
we wait until the scheduled time of the next request. This period allows adaptive generation protocol instances to generate new links with an updated internal state. If a request has not been completed after the scheduled time for the next request, we start to serve the next request immediately after the current is completed, leaving no wait time.
We record the latency of serving each request (as the number of time steps between submission of the request and its completion). 

At the scheduled time of a request, a path will be decided based on local best-effort routing~\cite{chakraborty2019distributed}, where nodes make path decisions based on existing entanglement links and network connectivity.
To complete the request, additional entanglement links may be needed. Thus, after a route has been selected, nodes along the path will prioritize the creation of on-demand entanglement links to service this request over continuous random generation. This may overwrite existing links.

Once the request queue has been serviced, we store the service latencies of each request $\ell(n)$ as a function of $n$ (the index of the request in the queue). 
We then reset the simulation by clearing the state of network nodes and setting the simulation time to 0. We repeat this trial $N_T$ times and average each $\ell(n)$ into the average latency of requests $\ell_{avg}(n)$. We also record an upper bound to the latency $\ell_{max}(n)$ as the ninety-fifth percentile of service latencies over all trials. Plots are made with a moving window average with window size 3.

For further details of our implementation, the source code is available in our GitHub repository~\cite{GitHub} for this project.


\subsection{Request Scenarios}
\label{sec:scenarios}
We consider multiple scenarios with increasing complexity:
\begin{enumerate}
    \item Scheduling a queue of equivalent requests (always between one pair of nodes) with a constant time interval between requests.
    \item Scheduling a queue of random requests (from the traffic matrix) with a constant time interval between requests.
    \item Varying the time between requests.
\end{enumerate}
The first scenario is the simplest and provides basic insight on the behavior of latencies.
The second takes interaction between different requests into consideration. Different requests might interfere with each other with regards to the adaptive protocol, and traffic load differences might further affect the performance of the adaptive protocol. The third scenario is the most realistic, and it requires the consideration of such effects as resource reservation and deadlock. In this paper, we focus on the first two scenarios and leave the third as future work.
\section{Simulation Setup}
\label{sec:setup}

\subsection{Adaptive Generation Protocol}
\label{sec:adaptive}
The adaptive generation protocol will continuously attempt to generate entanglement with adjacent nodes. It will randomly choose which adjacent node to attempt to generate entanglement with. Initially, the distribution of nodes will be uniform, such that a protocol instance on node $n$ may select each neighboring node $i$ with probability $p^{(i)}_0 = 1/d_n$ probability of being selected (where $d_n$ is the degree of node $n$).

\begin{wrapfigure}{R}{0.17\textwidth}
    \centering
    \includegraphics[width=0.15\textwidth]{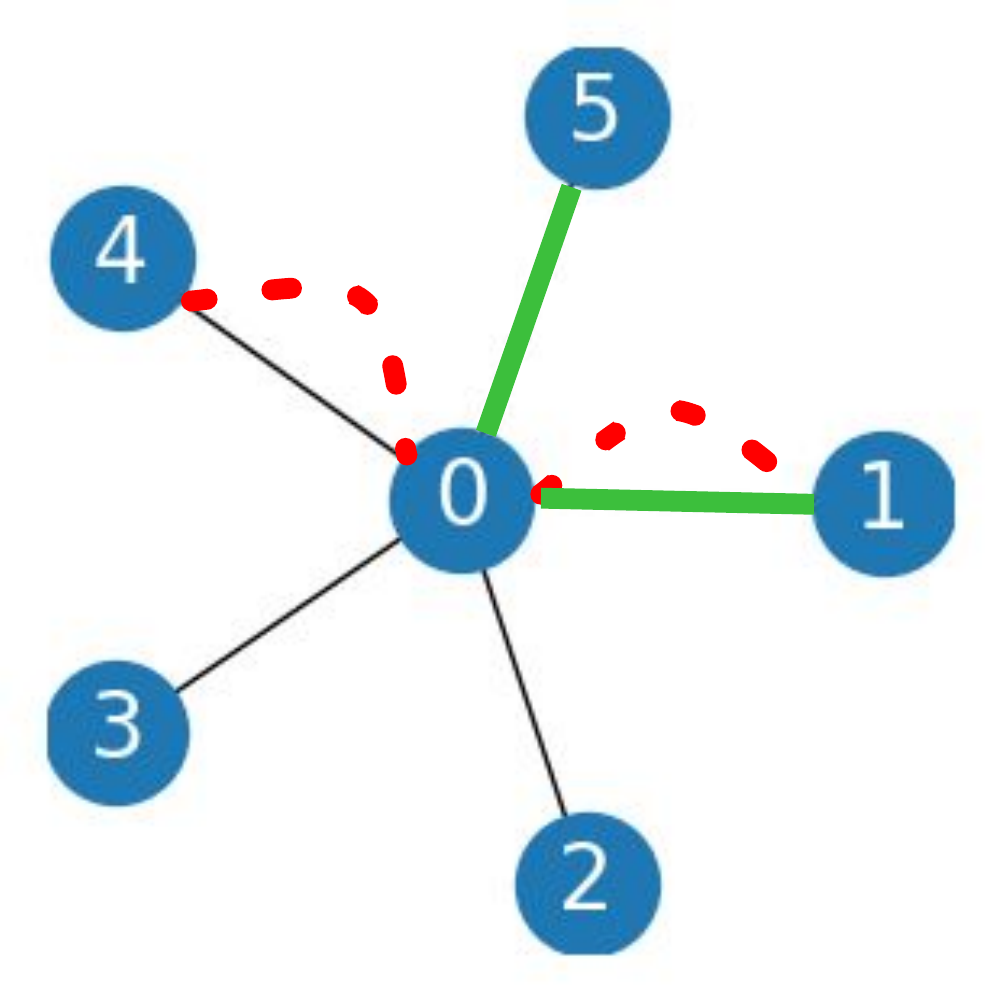}
    \caption{Neighbors of a central node $0$ with entanglement links (dotted red) and part of a found route (solid green). In this scenario, $S = \{(0,1)\}$ and $V = \{(0,5)\}$.}
    \label{fig:adaptive_neighbors}
\end{wrapfigure}{}

As a request is being serviced, the adaptive protocol will track the requested links and their overlap with the previously generated links; that is, it will track
\begin{itemize}
    \item The set of links $S$ which were pre-generated with direct neighbors and consumed in servicing the requests, and
    \item The set of links $V$ which were not pre-generated and had to be created on-demand to service the request.
\end{itemize}
The relation between these sets is shown in Figure~\ref{fig:adaptive_neighbors}. Once the request has been completed, the adaptive protocol will
\begin{itemize}
    \item Maintain the probability of selecting nodes corresponding to links in $S$,
    \item Increase the probability of selecting nodes corresponding to links in $V$, proportional to 1 minus the sum of probabilities for $S$ and $V$, and
    \item Decrease the probability of selecting nodes corresponding to unused links; that is, links that are not in $S$ or $V$.
\end{itemize}
Formally, this may be written as
\begin{gather}
    p_{t+1}^{(i)} = \begin{cases}
    p_{t}^{(i)} & \text{if } i\in S\\
    p_{t}^{(i)} + \frac{\alpha}{|V|}\left(1 - \sum_{j\in S\cup V}p_{t}^{(j)}\right) & \text{if } i\in V\\
    \frac{1}{d_n - |S\cup V|}\left(1 - \sum_{j\in S\cup V}p_{t+1}^{(j)}\right) & \text{otherwise }
    \end{cases}
\end{gather}
where we ensure that the sum of probabilities $\sum p^{(i)}$ equals 1 for all rounds. This protocol description introduces an additional simulation parameter $\alpha$, describing how quickly we change probabilities. This parameter has bounds $0 \leq \alpha \leq 1$ to keep probability distribution normalized. These probabilities will then be used until another request passes through the node.


\subsection{Network Setup}
\label{sec:net_setup}

Unless otherwise stated, the network parameters shown in Table \ref{tab:default_params} are used for each simulation. We test the performance of the adaptive protocol on 2 network topologies, namely an autonomous systems (AS) network and a 
single-bottleneck network. The AS-network is generated randomly using the networkx~\cite{networkx} library to mimic the design of the classical internet, providing interesting connectivity and paths.

\begin{table}[htbp]
\centering
\caption{Parameter List with Default Values.
}
\label{tab:default_params}
\begin{tabular}{|c|c|}
\hline
\textbf{Parameter} & \textbf{Value}\\ \hline
NET\_SIZE ($N$) & 10 \\ \hline
MEMO\_SIZE ($m$) & 5 \\ \hline
MEMO\_LIFETIME ($\tau_m$) & 1000 sim. time steps \\ \hline
ENTANGLEMENT\_GEN\_PROB ($p_e$) & 0.01 \\ \hline
ENTANGLEMENT\_SWAP\_PROB ($p_s$) & 1 \\ \hline
QUEUE\_INT ($\Delta T$) & 500 sim. time steps \\ \hline
\end{tabular}
\end{table}

We expect for near term implementation of local, metropolitan area quantum networks, the number of nodes will be a few tens or less, and quantum memories will remain an expensive resource for networking use so each node should also contain a few tens or less. Time parameters in the simulation are in units of simulation time step, whose corresponding real-world value is determined such that it is of the same order as entanglement generation time $T_{\mathrm{gen}}$ and swapping time $T_{\mathrm{swap}}$, i.e. we define a time $T_0 = O(T_{\mathrm{gen}})\approx O(T_{\mathrm{swap}})$ as time unit (the approximation is justified by the fact that both operations require information transmission over quantum and classical channels). The time for one trial of entanglement generation over a physical link is $L_0/c + T_{\mathrm{init}}$~\cite{sangouard2011quantum}, where $L_0$ is the length of the link, $c$ is the speed of light in a certain transmission medium, and $T_{\mathrm{init}}$ is the time to initialize a quantum memory. For example, a recently proposed protocol based on rare earth ions~\cite{asadi2020protocols} involves an initialization time of $T_{\mathrm{init}}\approx 85$~\textmu s, and given that $L_0\approx 10$~km, we have that $T_{\mathrm{gen}}$ is around a few hundred microseconds. For swapping, the usual procedure requires a Bell state measurement on the middle node and communication of the result to neighboring nodes, followed by local operations based on the measurement outcome. Therefore, the $T_{\mathrm{swap}}$ is also the sum of communication time and local operation time, being roughly the same order of $T_{\mathrm{gen}}$. If we take 1~ms as the time step interval, quantum memory lifetime $\tau_m = 1000$ corresponds to a lifetime of 1~s, which can be satisfied by e.g. trapped ions~\cite{wang2017single}, neutral atoms~\cite{wang2016single}, and rare earth ensembles~\cite{ranvcic2018coherence,zhong2015optically}. The unit swapping success probability refers to deterministic swapping, which was considered in previous simulation work~\cite{chakraborty2019distributed}. Entanglement generation success probability $p_e$ can be estimated as $(\eta_c\eta_t\eta_d)^2/2$~\cite{9492859}, where $\eta_c$ is coupling efficiency for light emitted by the quantum memory, $\eta_d$ is detector efficiency, and $\eta_t\approx e^{L/L_{\mathrm{att}}}$ is transmission efficiency, with characteristic attenuation length for optical fibers being $L_{\mathrm{att}}\approx 22$~km. Considering $L\approx 10$~km and with an optimistic expectation of high detection and optical coupling efficiencies, $p_e$ on the order of 0.1 might be possible for future implementations.

\section{Results and Discussion}
\label{sec:results}

In this section, we present the results of using our adaptive protocol on a few example network scenarios.

\subsection{AS Network Topology}
\label{sec:as_net}

\begin{wrapfigure}{R}{0.25\textwidth}
    \centering
    \includegraphics[width=0.24\textwidth]{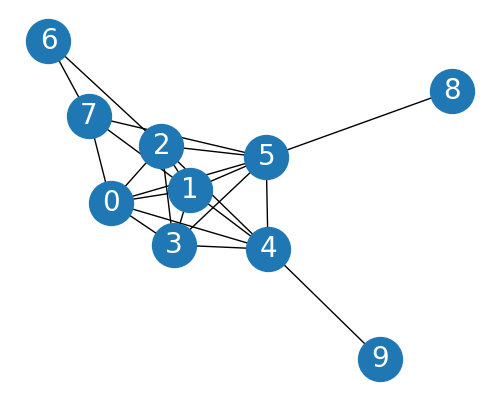}
    \caption{Network topology for a 10-node autonomous systems network.}
    \label{fig:10_node_topo}
\end{wrapfigure}

We first test the performance of our adaptive scheme on a simple, 10-node autonomous systems (AS) network. The topology for this setup is displayed in Figure~\ref{fig:10_node_topo}.


\begin{figure}[h]
    \centering
    \includegraphics[width=\linewidth]{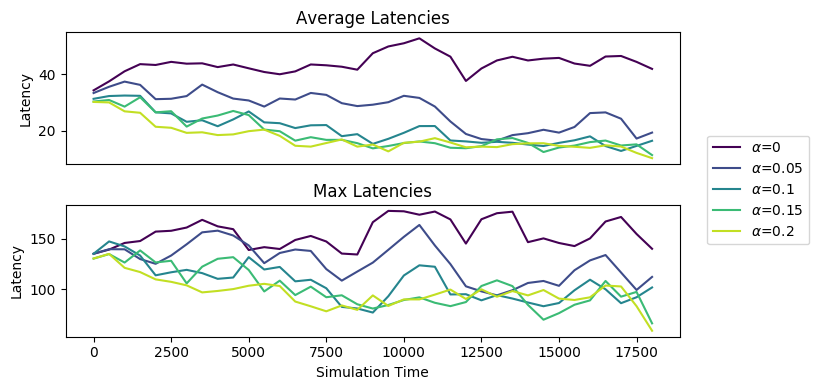}
    \caption{Service latency behavior for an AS-net topology with 10 nodes.}
    \label{fig:multi_small}
\end{figure}

First, we observe the average latency $\ell_{avg}(n)$ and max latency $\ell_{max}(n)$ for simulations with adaptive parameters ranging from $\alpha=0$ to $\alpha=0.2$. These results are shown in Figure~\ref{fig:multi_small}.
We note that the adaptive protocol is able to provide improved performance for scheduled requests, compared to the $\alpha=0$ case where there is no adaptation. Additionally, for higher $\alpha$, the adaptive protocol is able to  quickly converge to better performance; however, this has a limit of about $\alpha=0.1$ after which convergence does not improve significantly. 
This is because our adaptive protocol only updates the probability distributions on nodes in the found path, and thus nodes not in the path
will continue to generate entanglement with nodes in the path and consume memory resources.
{}

\begin{figure}[htbp]
    \centering
    \includegraphics[width=\linewidth]{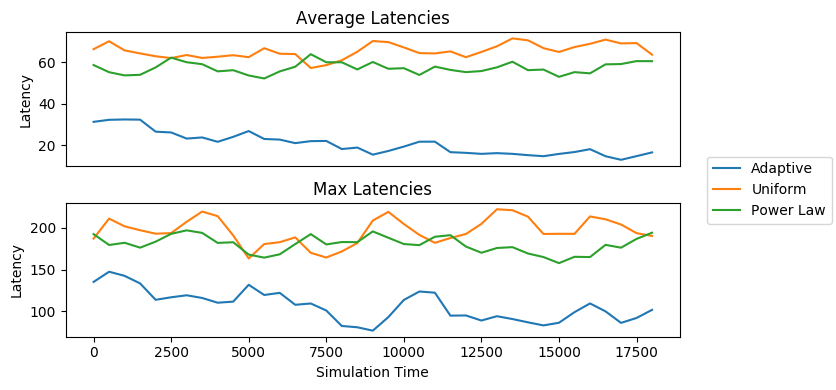}
    \caption{Service latency behavior for an AS-net topology with 10 nodes and various schemes for continuous entanglement generation.}
    \label{fig:compare_schemes}
\end{figure}

We also compare the $\alpha=0.1$ case to two fixed entanglement schemes mentioned in~\cite{chakraborty2019distributed}; these are
\begin{itemize}
    \item The uniform scheme, where each node may pick any node from across the network as an entanglement partner with uniform probability, and
    \item The power law scheme, where each node may pick any node from across the network as an entanglement partner with probability that decreases with distance.
\end{itemize}
Results for this comparison are shown in Figure~\ref{fig:compare_schemes}. We see that the performance of the adaptive scheme is better than both the power law and uniform schemes, even before the adaptive scheme has reached a convergence in performance. This is because the power law and uniform schemes allow choosing partners that are not direct neighbors for entanglement generation; when the partner is not a direct neighbor, to restrict entanglement generation attempts within one time step, we consider an attempt successful only if all needed elementary links are generated within that time step.
The success probability is then approximated as product of success probabilities for each elementary entanglement link. 
It is also for this reason that the power law scheme provides slightly better performance than the uniform scheme.
{}

\subsection{Single-Bottleneck Topology}

\begin{wrapfigure}{R}{0.45\columnwidth}
    \centering
    \includegraphics[width=0.45\columnwidth]{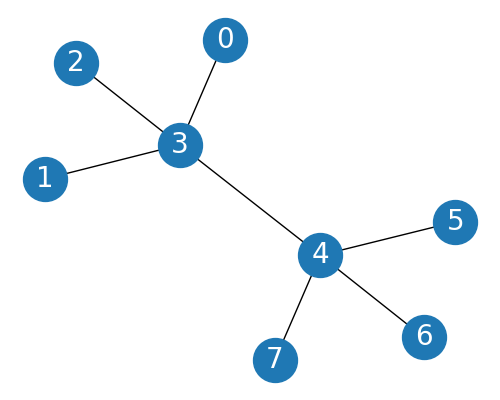}
    \caption{Network topology for an 8-node network with a bottleneck.}
    \label{fig:bottleneck_topo}
\end{wrapfigure}

In section~\ref{sec:as_net}, we focused on 
identical requests -- not to demonstrate our protocol's performance in a ``realistic'' scenario, but only to verify its concept. 
In this section, we consider a more realistic scenario where requests are randomly generated. Bottleneck structures are ubiquitous in various types of networks, and we envisage that they might also exist in quantum networks. Therefore, as a starting point of future exploration, we restrict ourselves to a simplified 8-node topology with a single bottleneck shown in Figure~\ref{fig:bottleneck_topo}.

The traffic matrix was then generated such that nodes will only request entanglement links with nodes on the other end of the bottleneck. This matrix is available on our GitHub page~\cite{GitHub}. We present results for a specific choice of parameters: adaptive weight $\alpha=0.05$ and request interval $\Delta T=200$. Note that our simulation is small scale and it is not hard to simulate for a different parameter choice. 

We consider two scenarios: the case of equal memory allocation to each node, and unequal memory allocation based on the degree of nodes. We expect decreased latencies from the unequal allocation, analogous to a classical network. We justify this by considering that during the random generation phase, each edge node has only one direct neighbor (a bottleneck node) to attempt entanglement generation with. Thus, their entanglement attempts will occupy bottleneck node memories. For equal memory allocation, this will exhaust the memory resources of the bottleneck nodes, preventing entanglement links across the bottleneck. With unequal allocation, edge node memories will be exhausted faster, leaving bottleneck nodes memories to generate entanglement links across the bottleneck with (which are necessary for any request). This reduces the probability that a bottleneck entanglement link is missing when serving requests, thus reducing latency.

\subsubsection{Equal Memory Number per Node}
\begin{figure}[htbp]
    \centering
    \includegraphics[width=0.9\linewidth]{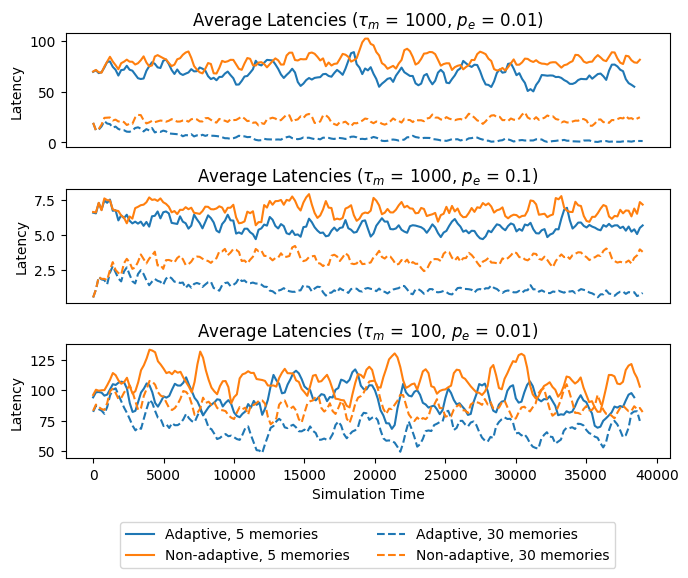}
    \caption{Service latency behavior for a bottleneck topology with various parameter choices and continuous entanglement generation schemes.}
    \label{fig:bottleneck}
\end{figure}

We first consider the scenario where every node has the same number of quantum memories. Based on the expectation that near-term networks will have a small number of quantum memories per node, we arbitrarily choose two different quantities, 5 and 30 memories, 
to explore the effect of resource scarcity on the protocol performance, and we also consider different quantum memory lifetimes and entanglement generation probabilities. We compare our adaptive scheme to a non-adaptive scheme (i.e. with $\alpha=0$), with average latencies shown in Figure~\ref{fig:bottleneck}.

{}

First, we note that the adaptive protocol gives lower latencies for later requests (even with randomized requests). For different sets of parameters, more quantum memories per node lead to lower average latencies. Additionally, we observe that by increasing quantum memory lifetime $\tau_m$ and entanglement generation probability $p_e$ the average latencies decrease. These results are in accordance with the intuition that more resources of higher quality will lead to better performance.

\subsubsection{Quantum Memory Allocation}
\begin{figure}[htbp]
    \centering
    \includegraphics[width=0.9\linewidth]{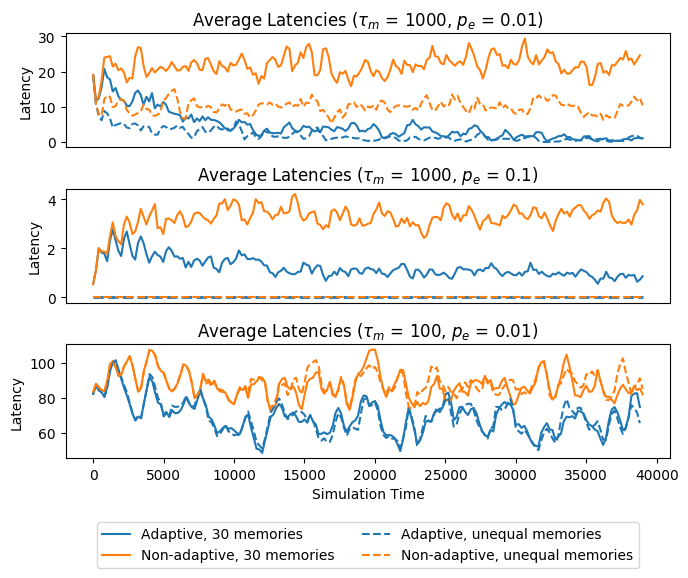}
    \caption{Latency behavior for a bottleneck topology with unequal memory count and continuous schemes.} 
    \label{fig:bottleneck_multi_memo}
\end{figure}

In the network topology from Figure~\ref{fig:bottleneck_topo}, the physical neighbors and traffic are not uniform. Therefore, it is natural to conjecture that quantum memory allocation on nodes might also influence network behavior. We explore the scenario where edge nodes (nodes 0-2, 5-7 from Figure~\ref{fig:bottleneck_topo}) are assigned 5 memories and bottleneck nodes (nodes 3 and 4) are assigned 30. In Figure~\ref{fig:bottleneck_multi_memo}, we compare this new allocation to the case where every node is assigned 30 memories.

{}

We observe significant improvement with unequal quantum memory allocation over equal allocation of 30 memories for $(\tau_m=1000, p_e=0.01)$ and $(\tau_m=1000, p_e=0.1)$. We note that unequal allocation leads to lower latencies for both adaptive and non-adaptive protocols (even approaching zero for $(\tau_m=1000, p_e=0.1)$). We additionally observe that the adaptive protocol still provides a decrease in latency over a non-adaptive protocol for unequal memory allocation (except for the $(\tau_m=1000, p_e=0.1)$ case, where latencies are already zero for both schemes).

Finally, we analyze why zero latency for $(\tau_m=1000, p_e=0.1)$ is possible in our simulation. Given $p_e=0.1$, on average a single entanglement link can be generated after 10 attempts.
With limited memory $m = 5$ for edge nodes, the expected number of entanglement links $E(L)$ is thus capped at $5$ for $\Delta T > 50$ (where a queueing interval $\Delta T = 200$ was used for the simulation). 
Under the most extreme scenario where both bottleneck nodes only attempt to generate entanglement with each other, within 50 time steps on average 10 memories will be occupied by entanglement links over the bottleneck. This leaves enough excess memories for entanglement with edge nodes. Thus, within $\Delta T = 200$ all elementary entanglement links will be generated for this specific topology and memory allocation.
Moreover, with $\tau_m=1000$, pre-generated entanglement links have a long lifetime and may be used to service future requests. As a result, required entanglement links for any request may be generated before the request is scheduled, leading to a latency of zero.
On the contrary, if we consider a smaller quantum memory lifetime (e.g. in our simulation $\tau_m=100<\Delta T=200$), the pre-generated entanglement links are prone to expiration before the submission of later requests. This is true regardless of memory exhaustion. Therefore, there will likely exist links which must be generated on demand. This explains the performance similarity between the two allocation schemes for $(\tau_m=100,p_e=0.01)$.

These results confirm our expectation that simply adding more resources will not always lead to improved performance for quantum networks, as is the case with classical networks.
\section{Conclusion}
\label{sec:conclusion}

We proposed an adaptive protocol that uses pre-generated entanglement links between neighboring nodes to reduce service latencies for entanglement requests between arbitrary nodes in a quantum network.
We also provided a simple but effective framework for qualitative analysis of request serving behavior. We then built a time-step simulator for generation of entanglement links and request servicing on a quantum network. Our simulation results show the effectiveness of our adaptive protocol for homogeneous request queues and certain network parameters, as compared to continuous entanglement generation schemes relying on static distributions. 
We also studied latency reduction for randomized requests on a single-bottleneck topology. Moreover, the difference in protocol performance for different methods of assigning quantum memories to nodes confirms that sophisticated resource allocation is required for optimizing quantum network performance. We leave further study of the network parameter space, more detailed network simulation, and inclusion of quantum effects within our protocol for future work.

\section*{Acknowledgment}
We thank Manish Kumar Singh for useful insights to improve our simulation.
A. K. and A. Z. thank Ben Y. Zhao, from whose course this project originated.

\bibliographystyle{IEEEtran}
\bibliography{references}

\end{document}